\documentstyle[12pt]{elsart}
\begin{document}
\addtolength{\topmargin}{50pt}
\def\today{24 March 2000}

\def\beq{\begin{eqnarray}}
\def\eeq{\end{eqnarray}}



\hrule

\noindent{\small\sc An Essay Written for the 51st Essay Competition of the
Gravity Research Foundation. Awarded Fifth Prize. To appear by invitation 
in Gen. Rel. Grav.} 
\vskip.5cm
\hrule

\vskip 1cm
\noindent
{\Large\sc Probing Quantum Violations of the Equivalence Principle}
\vskip 1cm
\noindent
{\sc G.\ Z.\ Adunas, E.\ Rodriguez-Milla, and D.\ V.\ Ahluwalia} 
\footnote{
{\sc Correspondence:}
ISGBG, Ap. Pos. C-600,
Escuela de Fisica de la UAZ, Zacatecas,
ZAC 98068, Mexico.
{\sc E-mail:} ahluwalia@phases.reduaz.mx. {\sc URL,} http://phases.reduaz.mx }

\vskip 1.5cm


{\sc Abstract}

The joint realm of quantum mechanics and the general-relativistic 
description of gravitation is becoming increasingly accessible to 
terrestrial experiments and observations. In this essay we study 
the emerging 
indications of the violation of equivalence principle (VEP). 
While the solar neutrino anomaly may find its natural 
explanation in a VEP, the statistically significant discrepancy observed 
in the gravitationally induced phases of neutron interferometry 
seems to be the first indication of a VEP.
However, such a view would seem immediately challenged by
the atomic interferometry results.  
The latter experiments see no indications of VEP, in apparent contradiction
to the neutron interferometry results.
Here we present arguments that support the view that these, and related
torsion pendulum experiments, probe different aspects of gravity; and that
current experimental techniques, when coupled to the solar-neutrino data, may
be able to explore quantum mechanically induced violations
of the equivalence principle. We predict quantum violation of the 
equivalence principle (qVEP) for next generation of atomic interferometry 
experiments. The prediction entails comparing free fall of two different
linear superpositions of Cesium atomic states.




\newpage
\noindent
{\sc I. Introduction}

Principle of equivalence is the empirical foundation of the 
general-relativistic description of gravitation. While all classical
experiments confirm the equality of the inertial ($m_i$) and 
gravitational masses ($m_g$), there has arisen an experimental 
hint which motivates us to review the operational basis of this 
equality in the quantum realm. 

More precisely, the ratio $m_g/m_i$ is empirically found to be 
composition-independent, and this translates into the 
composition-independent equality of gravitationally-induced 
accelerations in terrestrial laboratories. 

The classical tests of the equivalence principle, based
on sophisticated torsion pendulums, reveal no violations of the equivalence
principle. Quantitatively, in an experiment published in December 1999
\cite{tp}, the observed differential accelerations
of the Copper (Cu)  and Lead (Pb) test bodies toward
a $3$ ton $^{238}U$ attractor was found to be
\beq
{a}_{Cu}- a_{Pb} = \left(1.0\pm 2.8\right)\times 10^{-13}\,\,\mbox{cm/s}^2
\eeq
as compared to the corresponding gravitational acceleration of $9.2\times
10^{-5}\,\,\mbox{cm/s}^2$. While this test, in its essence, is purely
classical, and was designed to probe $\exp(-r/\lambda)$ type deviations
in the Newtonian limit of general relativity, another experiment
(reported in August 1999 \cite{chuetal}) compared gravitationally
induced accelerations of a classical
object (a macroscopic glass object) with that of a 
quantum object (a Cesium atom in a linear superposition of two 
different energy eigenstates). 
Denoting by $a_G$ the acceleration of the glass
object, and by  $a_{Ce}$, the acceleration of the indicated
Cesium atom; $a_G$ and $a_{Ce}$ were found identical to
$7$ parts in $10^{9}$. 

In the last twenty five years the Colella-Overhauser-Werner 
(COW) class of experiments \cite{COW1975} have become more sophisticated.
The latest (September 1997
\cite{COW1997}) neutron interferometry experiments report a 
statistically significant discrepancy between 
the experiment and theory, and it has been suspected (by the experimenters)
to carry any of the following two sources: (a) some systematic error in the
measurements [to be called type-A], and
(b) ``they (the discrepancy) may also represent a difference 
between the ways in which gravity acts in classical and quantum mechanics,''
to directly quote the authors \cite{COW1997}. The gravitationally induced
differences in phases that the experiment probed depend on the
following combination of the inertial and gravitational masses
\beq
u=  \frac{m_i \,m_g \,g}{2\pi\hbar^2}\times 
\mbox{a geometrical factor} 
\eeq
where $g$ is the acceleration due gravity.
 Assuming the equality of the inertial and the gravitational masses for
neutron, $m_i=m_g$,
the experiment found this phase factor to be about $1\%$ lower than predicted:
\beq
\frac{\left(u\right)_{expt.} - \left(u
\right)_{m_i=m_g}}{    \left(u
\right)_{m_i=m_g}}
=\cases{ -(1.5 \pm 0.12)\times 10^{-2},   \cr
-(0.8 \pm 0.11)\times 10^{-2} \cr }
\eeq
where the top value corresponds to the skew-symmetric 
interferometer while the bottom value
is associated with the symmetric interferometer. This type of anomaly shall
be called type-B anomaly.

It would thus appear that
the latest neutron interferometry experiments 
are in conflict with the more precise tests of the equivalence principle
conducted via atomic interferometry, and with those
based on torsion pendulum. 
In fact, authors of Ref. \cite{chuetal} write ``we may conclude that there are
aspects of neutron interferometry that are not well understood.''
While that may be so in part or totality, 
here we shall argue that the answer is not necessarily so simple. 
In essence, we support the view  
that each of these experiments probes a different 
aspect of the equivalence
principle. Furthermore, we predict that the atomic interferometry experiments,
as the precision improves further, should begin to see a violation of the
equivalence principle.

We deliberately enter details, which may appear common sense, in order
to raise relevant questions that may otherwise escape our attention. 
We also add that very recently a new type of neutron interferometer
has emerged \cite{new} and it should be able to provide a conclusive
statement on the nature of the anomaly observed in neutron interferometry.
In particular these experiments have the potential to rule out type-B
anomaly at about $0.5\%$ level.\footnote{The
experiment described in Ref. \cite{new} was published in February 2000,
and did not come to our attention till after the submission of 
this {\em Essay\/} to the {\em Gravity Research Foundation.\/} Any reference
to \cite{new} must be considered as an addenda to the original  
 {\em Essay.\/}}
However, it is precisely this ability of the
experiments to probe type-B anomaly that intrigues us here. One should
in fact take a conservative view, as supported by the preliminary
results of \cite{new}, that experiments of the type described in 
Ref. \cite{new} will indeed fully rule out the anomaly 
found in Ref. \cite{COW1997}. But, at the same time we will take the argued 
view that the inertial and gravitational masses are operationally independent
objects, and that neutron interferometers remain a powerful tool to
experimentally study equality of the inertial and gravitational masses. 
With this view we shall examine various experimental settings 
to study quantum violations of the equivalence principle.

{\sc II. Quantum Violations  
of the Equivalence Principle: The discrepancy in Neutron Interferometry
Experiments}

Unlike an electron, which is the active player in the atomic interferometry,
a neutron is a complicated
object containing the $udd$ quark configurations coupled
to $\bar q q$ sea quarks and the gluonic degrees of freedom. These degrees of 
freedom, in terms of their spatial distribution, spend a fraction of their
time in the classically forbidden region. To gain physical insight into
the question of neutron-Earth 
gravitational interaction\footnote{The latter 
in turn carrying more than half its mass in 
neutronic matter.} 
we note that the 
baryonic spectra reveals a series of excited neutron states,
and that these states are roughly equally spaced. Thus in 
``back of the envelope spirit''
we can treat neutron as  a ground state of an harmonic
oscillator. This would not alter our general qualitative 
results in any significant manner.

For the ground state of an harmonic 
oscillator\footnote{We restrict to a one-dimensional non-relativistic case
because this serves as a good representative example to study
classically forbidden region.} we evaluate the kinetic and
potential energy contributions from the classically allowed region (CAR), 
and from the classically forbidden region (CFR). We find that for the ground
state of an harmonic oscillator,  
the CAR contributions to the potential
and kinetic energy are, respectively: 
\beq
&&
E^{\sc CAR}_{0,P}=
\left(\frac{\mbox{erf}(1)}{2}-\frac{1}{\e\sqrt{\pi}}\right)
\frac{\hbar\omega}{2}\label{pcar}
\\
&&
E^{\sc CAR}_{0,K}=
\left(\frac{\mbox{erf}(1)}{2}+\frac{1}{\e\sqrt{\pi}}\right)
\frac{\hbar\omega}{2}\label{kcar}
\eeq
Similarly, the contributions from the CFR
read:
\beq
&&E^{\sc CFR}_{0,P}=\left(
\frac{1}{2}
+
\frac{1}{\e\,\sqrt{\pi}}
-
\frac{\mbox{erf}(1)}{2}
\right)\frac{\hbar\omega}{2}\label{pcfr}\\
&& E^{\sc CFR}_{0,K}= 
\left(
\frac{1}{2}
-
\frac{1}{\e\,\sqrt{\pi}}
-
\frac{\mbox{erf}(1)}{2}
\right)\frac{\hbar\omega}{2}\label{kcfr}
\eeq
In these expressions ``$\e$'' 
represents the natural base of logarithms (and
roughly equals $2.718$), $\mbox{erf}(x)$ is the standard error function,
and other
symbols have their usual meaning. The total ground state 
energy, $E_0=(1/2) \hbar\omega$, carries the following proportions:
$\mbox{erf}(1)\times E_0$ from the CAR, and $\left[1-\mbox{erf}
(1)\right]\times E_0$ from
the CFR. 
Since $\left[1-\mbox{erf}(1)\right] \approx 0.16$, 
roughly sixteen 
percept of the ground state energy is contributed by the CFR. 
Moreover, as is not unexpected,
the total contribution to the kinetic energy from CFR is negative definite
and equals $\approx - \,0.13 E_0$.\footnote{Note that CAR contribution 
combined with CFR contribution adds to $(1/4)\hbar\omega$ for kinetic 
as well as
the potential energy.}
 In fact,
the kinetic energy density\footnote{In the  
expression below $\psi_0(x)$ represents the ground state 
wave function for 1-D harmonic oscillator.
Also note that $\int_{-\infty}^\infty \rho_{0,K}(x) dx = (1/4)\hbar\omega$, 
consistent with results given in Eqs. (\ref{kcar}) and (\ref{kcfr}).}
\vbox{
\beq
\rho_{0,K}(x)&&\equiv
\psi_0^\ast(x) \left[-\frac{\hbar^2}{2 m}
\frac{\partial^2}{\partial x^2}\right]
\psi_0(x)\nonumber\\
&& = \frac{1}{2}\sqrt{\frac{m\omega}{\pi\hbar}}
\left(\hbar\omega-m\omega^2 x^2\right)
\exp\left[-\,\frac{m \omega x^2}{\hbar}\right]
\eeq
}
is positive definite only for CAR. It is negative definite for the entire
CFR as a simple graphical analysis reveals. 
This is not surprising at all since momentum
is formally imaginary in {\em all\/} CFRs. 

The point to be made now is as follows.
Generally, it is argued that this formally imaginary momentum does
not create any paradoxical situation 
because any attempt to confine the system to
a classically forbidden region and to {\em measure\/} 
its momentum always imparts
the system enough energy to destroy the very system one wishes to observe.
It is only in the last decade 
that the CFRs have becomes experimentally accessible.
Recent experiments on tunneling times indicate that the standard wisdom
may not be the entire story\cite{fs1,fs2,fs3,fs4,fs5,fs6}. 
Specifically, experiments such as described in Refs. \cite{fs1,fs4}
do not probe the {\em spatial extent\/} of the classically forbidden
region. Instead, they simply assure that a photon (or, some similar
probe) encountered a CFR on it way from the 
source to the detector.
In the context of neutron interferometry,
the general-relativistic gravity not only probes the total energy, 
as encoded in the  the time-time component of the energy-momentum tensor,
$\tau^{\mu\nu}(x)$,
of the {\em test\/} particle, but it is also sensitive to the
energy currents (such as those represented
by momentum flux coming from CFR and
encoded in the time-space part of the 
$\tau^{\mu\nu}(x)$.\footnote{The initial theoretical lesson learned from
the neutron interferometry was that there exist quantum gravitational effects
that depend on the test-particle mass. This circumstance arises
from the fact that the test-particle mass does not cancel out
from the quantum equations of motion despite the equality of the inertial
and gravitational masses. What we are now proposing is that since the 
``gravitational charge'' is related to the $\tau^{\mu\nu}(x)$, 
one needs to look
beyond the time-time component of  $\tau^{\mu\nu}(x)$ in neutron interferometry
and consider the neutron  $\tau^{\mu\nu}(x)$ as its gravitational charge. 
For extended test particles, such as neutrons, this may carry non-trivial
physical consequences. The source-$T^{\mu\nu}(x)$ already has a  starring role
in the theory of general relativity. By fully 
extending that role to test particles
we shall not only introduce a ``source--test-particle'' 
symmetry but at the same time open a way to experimentally study it.}
For this reason,
and because these enter phases in quantum mechanical evolution of a quantum
systems, CFR may affect gravitational evolutions of a system. Since neutron
(and Earth) must carry CFR contributions, the
discrepancy observed in the gravitationally induced phases in 
in the latest neutron interferometry experiments may be  probing this 
anomalous neutron-Earth interaction. 
In case, the present discrepancy is ruled out by the new type 
of neutron interferometer \cite{new}, one should very much encourage
development of still more ingenious neutron interferometers with
accuracies far beyond a fraction of $1\%$. In such experiments one may
indeed see a violation of the equivalence principle. This violation
may arise from the CFR contributions to source and test-particle
energy-momentum tensors, or may arise from the influence of an 
essentially constant gravitational potential
due to the cosmic distribution of matter.
The latter has no local
physical observability in a framework in which the equivalence principle
is respected, but once one allows for a VEP the indicated potential
carries deep physical significance \cite{qVEP}.

The gravitational role
played by the energy-momentum tensor is far more intricate in a 
general-relativistic description of gravity. It may become even more
intricate in a quantum mechanical settings as already 
suspected by Littrel, Allman, and Werner  
\cite{COW1997}, and even if the anomaly
they published turns out to be of type-A.

Already in 1975 \cite{COW1975}, it was {\em experimentally\/} established 
that the equality of gravitational and inertial masses does not imply
that the mass of a test particle shall drop out of the 
quantum-gravity equations of motion.
The latest neutron interferometry results suggest that
these experiments may have evolved 
to an extent that some of them are probing the CFRs. 
The effects of the latter identically
vanish if one simply describes  the test particle by 
expectation value of the test-particle $\tau^{\mu\nu}(x)$.

\noindent
{\sc III. Quantum Violations  
Of the Equivalence Principle: Atomic Interferometry}

In a framework in which the inertial and gravitational masses are considered
operationally independent objects, it is evident that
one should expect a tiny violation of the equivalence principle
in the quantum regime. Since every physical
system carries an inherent energy uncertainty determined by
$\Delta E \Delta t \sim \hbar$, its inertial and
gravitational properties must carry unavoidable 
fluctuations. In particular,
these fluctuations affect the equality of the inertial and gravitational 
masses,
and may even emerge as violation of the 
equivalence principle.
However, as time of observations takes on macroscopic values these fluctuations
become vanishingly small and only very clever experiments, perhaps along the
lines suggested by Amelino-Camelia (in an entirely different context
\cite{gacN}), could 
be hoped to probe these fluctuations in the equality of the inertial and
gravitational masses. 

Here, we take a far more readily accessible experimental situation and
study a possible violation, and associated fluctuations, 
of the equivalence principle.
In effect, we choose a system for which certain quantum fluctuations evolve 
coherently.

To model an experimental set up, such as that used in the 
atomic interferometry experiments \cite{chuetal}, 
consider two ``flavors'' of Cesium atoms:
\beq
\left[
\matrix{\vert ^{\alpha}Ce\rangle_\xi\cr
        \vert ^{\beta}Ce\rangle_\xi}\right]
= \left[\matrix{\cos(\xi) & \sin(\xi) \cr
	-\sin(\xi) & \cos(\xi)}\right]
\,
\left[
\matrix{\vert ^{E_1}Ce\rangle\cr
        \vert ^{E_2}Ce\rangle}\right]
\eeq 
Here, $\vert ^{E_1}Ce\rangle$ and $\vert ^{E_2}Ce\rangle$ represent two
different {\em energy\/} eigenstates of the Cesium atom. 
The ``flavor'' states, $\vert ^{\alpha}Ce\rangle_\xi$ and 
$\vert ^{\beta}Ce\rangle_\xi$, 
are linear superposition of the energy eigenstates. These 
are characterized by the flavor indices $\{\alpha, \beta\}$,
and by the mixing angle $\xi$. In a given 
gravitational environment these 
flavors oscillate from one flavor to another as is now well understood
\cite{ws6,ws1,ws2,ws3,ws4,ws5}.
The oscillation of the flavors provides a flavor-oscillation clock, and
the flavor-oscillation clocks red-shift as required by the theory of general 
relativity. Here we shall concentrate on an entirely different issue,
and exploit the fact that the flavor-oscillations carry the fluctuations
$\Delta E \Delta t \sim \hbar$ in a coherent manner --- the inherent 
energy-uncertainty associated with the flavor states is simply 
related to the inverse of time period of the flavor-oscillation. However, 
this evolution happens in coherent manner and does not suffer from randomness
often associated with the constraint $\Delta E \Delta t \sim \hbar$.

Having emphasized that we are modeling an experimentally accessible situation
that can be realized at the Stanford laboratory of Steven Chu and, 
possibly also
at the Wineland-Itano's group at NIST, we now introduce a simplified notation:
\beq
&&\vert\psi_{\ell\xi}\rangle = \vert^\ell Ce\rangle_\xi,
\quad\ell=\alpha,\beta\\
&&\vert\varphi_{E_j}\rangle=\vert^{E_j} Ce\rangle,\quad j=1,2
\eeq

We assume that the states  
$\vert\varphi_{E_j}\rangle$ are,
in comparison to the time of observation, long lived. The ``free'' 
fall experiment that we consider assumes, for simplicity, that flavors 
do not evolve significantly during their ``free'' fall from the source to
the detector. Relaxing these assumptions to suit a given experimental
situation should pose no technical or conceptual problem.

The flavor states of the Cesium atoms carry an inherent uncertainty
in their energy
\beq
\Delta E_{\ell\xi} = 
\sqrt{\langle\psi_{\ell\xi}\vert H^2\vert\psi_{\ell\xi}\rangle
-\langle\psi_{\ell\xi}\vert H\vert\psi_{\ell\xi}\rangle^2}
\label{uncer}
\eeq
where
\beq
 H \vert \varphi_{E_j}\rangle
= E_j \vert  \varphi_{E_j}\rangle, \quad j=1,2
\eeq
The uncertainty (\ref{uncer}) is what a set of large number of 
energy-measuring experiments, on identically prepared flavor states, 
would yield. While it has the same structural form as that of a statistical 
error, it is not associated with the measuring devices and does not go 
to zero as $1/\sqrt{N}$ (where $N$ is the number of measurements). 
It is an irreducible quantum uncertainty that characterizes a given 
flavor eigenstate.\footnote{One of us (DVA) 
thanks Mariana  Kirchbach for several long discussions
on this point.}
For this reason, the equality of the gravitational and inertial masses 
for flavor states carries
an inherent violation of the equivalence principle. The associated
quantum violation in the equivalence principle (qVEP) can be characterized
by the fractional accuracy
\beq
f_{\ell\xi}= \frac{\Delta E_{\ell\xi}}{\langle E_{\ell\xi}\rangle}
\eeq
where $\langle E_{\ell\xi}\rangle \equiv\langle\psi_{\ell\xi}\vert H\vert 
\psi_{\ell\xi}\rangle$

Thus, if we study two sets of Cesium atoms, 
with flavors characterized by angles $\xi_1$ 
and $\xi_2$, then qVEP predicts a {\em difference in the
spread in their accelerations\/}
(as observed in identical  free fall experiments by a stationary
observer on Earth) to be:
\beq
 \vert \Delta a_{\ell\xi_2}\vert
 -  
\vert \Delta a_{\ell\xi_1}\vert = 
\left\{\left\vert
\frac{\sin(2\xi_2)}{\langle E_{\ell\xi_2}\rangle}
\right\vert
-
\left\vert
\frac{\sin(2\xi_1)}
{\langle E_{\ell\xi_1}\rangle} \right\vert
\right\}
\frac{\delta E}{2},
\eeq
where $\delta E \equiv E_2-E_1$.
For flavor states of Cesium atoms prepared with $\delta E 
\approx 1 \,\,\mbox{eV}$, this difference is of the order of a few parts
in $10^{12}$, and should be observable in refined versions of experiment
reported in Ref. \cite{chuetal}. How difficult this refinement in 
techniques at Stanford and NIST would be is not fully known to us.
However, the extraordinary accuracy in similar experiments and the 
already achieved absolute uncertainty of $\Delta g/g\approx 3\times 10^{-9}$,
representing a million fold increase compared with previous experiments,
makes us cautiously optimistic about observing qVEP in atomic 
interferometry experiments pioneered  by the group of Steven Chu at Stanford.

For additional discussion of this experiment, see Ref. \cite{qVEP}
--- {\em cf.\/} Ref. {\cite{vo}.

{\sc IV. Quantum Violations  
of the Equivalence Principle: Solar Neutrino Anomaly}

It has long been conjectured that the solar neutrino anomaly may be related
to a flavor-dependent violation of the equivalence principle.
The suggestion first came from Gasperini \cite{mg1,mg2}.
The argument presented above can -- with appropriate generalization, 
and on interpreting
the flavor index $\ell$ to represent the three neutrino flavors ($\nu_e$,
$\nu_\mu$ and $\nu_\tau$) -- be extended to neutrino oscillations.
Here we simply provide an outline of this argument and show how
a quantum violation 
of the equivalence principle naturally arises.
 
To estimate the qVEP effects it suffices to restrict
to a two-state neutrino oscillation framework. A simple calculation shows
that the difference in fractional measure of qVEP turns out to be 
exceedingly small:
\beq
\Delta f_{\ell\ell'}\equiv f_\ell -f_{\ell'}  = 6.25 \times 10^{-26} \left[
\frac{(\Delta m^2)^2}{\mbox{eV}^4}\right]
\left[\frac{\mbox{MeV}^4}{E^4}\right] \sin(4\xi_V)  
\eeq
where $\ell$ (say, $\nu_e$) 
and $\ell'$ (say, $\nu_\mu$) refer to two different neutrino flavors, and
$\xi_V$ is the vacuum mixing angle between the underlying mass
eigenstates (whose superposition leads to different flavors of neutrinos).
The difference in the squares of the underlying mass eigenstates, 
$m_2^2-m_1^2$, has been represented  by $\Delta m^2$;  and $E$ is the 
expectation value of the neutrino energy. For the solar neutrinos 
the existing data spans the approximate range $ 0.2\,\,\mbox{MeV} 
\le E \le 20\,\,\mbox{MeV}$ in energy.

Interestingly,
following standard arguments that yield an oscillation length from
a flavor-dependent violation of the equivalence principle \cite{solarVEP1},
 this is precisely this {\em smallness\/} of $\Delta f_{\ell\ell'}$  
that gives rise to a {\em large\/}
oscillation length for solar neutrinos that compares well with the 
Earth-Sun distance. Scenarios of the violation of equivalence principle
are currently under active investigation 
\cite{solarVEP1,solarVEP2,solarVEP3,solarVEP4} 
to explain the
solar neutrino anomaly (also see, Refs. \cite{srp1,srp3}). 
 They have the advantage that none of the three
separate evidences (the atmospheric, the accelerator, and the solar)
for neutrino oscillations need to
be ignored to make a consistent fit to all existing data. The qVEP induced
oscillation length not only matches the Earth-Sun distances, but it also
differs from the standard scenario of Gasperini where one assumes a
energy-independent violation of the equivalence principle. The qVEP
induced oscillation length carries a $E^3$ energy dependence, and would be
clearly distinguishable from Gasperini's conjecture as more data becomes
available, and if (as is true for all such analysis) the constant 
gravitational potential due to local supercluster of galaxies turns out to
have the expected value. It is to be noted that such a gravitational potential
carries little significance for planetary orbits because it is essentially
constant over the solar system. However, it turns out to be important for the
gravitationally induced phases that determine the qVEP induced effects (or 
even those that arise from the VEP conjectured by Gasperini).

For a more complete discussion of this issue, see Ref. \cite{qVEP}. There,
certain issues arising from the interplay of the kinematically induced
oscillation length and a qVEP-induced oscillation length are discussed
at the needed length.

\noindent
{\sc V. Concluding Remarks}

In the last decade the propagation through tunneling regions has 
probed classically forbidden regions with some dramatic, and still
controversial, results. Paralleling this development, 
neutron interferometry,
we suggest, may be probing, or may become capable of 
probing, the classically forbidden region
where the momentum-density is pure imaginary, and kinetic
energy density negative definite. That matter inside a
neutron must exist in classically forbidden regions
is a general expectation of all quantum mechanical considerations and
is independent of specific models in its qualitative aspects.

However, there is more than one aspect of quantum mechanical structure
that requires a deeper study in the context of gravity. This has been
made abundantly clear in recent years. Fortunately, many of these 
aspects, such qVEP, may be studied in atomic interferometry and may
lie at the heart of the explanation of the solar neutrino anomaly.
If this was proven, then in the spirit of the concluding remarks 
in Ref. \cite{dvaN}, we may once again say that while Planck scale 
physics seems so remote it does not make quantum gravity a science 
where humans cannot venture to probe her secrets.

\noindent
{\sc Acknowledgements} 

It is my (DVA) 
pleasure to thank Giovanni Amelino-Camelia
for our continuing correspondence on matters of quantum gravity 
phenomenology, Maurizio Gasperini for a series of e-mail exchanges on
VEP and qVEP, Carlo Rovelli for raising a series of insightful questions 
via an extended e-mail exchange, Sam Werner for updates on the COW 
experiments, and G. van der Zouw for providing Ref. \cite{new}.

During some of these discussions a point of nomenclature has arisen.
It is this that usually
one considers quantum gravity as a subject that
studies quantum aspects of gravity. What we study here is behavior of
quantum objects in a classical gravity background. In the framework
of this essay, a detection of 
VEP/qVEP may warrant reconsideration of the  classical structure of gravity.  
That is, a discovery of VEP/qVEP 
may alter the very meaning of gravity in ``quantum gravity.'' 
For this reason, the proposals discussed here, and elsewhere \cite{qVEP},
have been dubbed ``quantum gravity phenomenology'' --- {\em cf.\/} footnote
\#7.



\end{document}